\documentclass[twocolumn,amsmath,amssymb,floatfix]{revtex4}

\usepackage{graphicx}
\usepackage{dcolumn}
\usepackage{bm}
\usepackage{color}
\usepackage{subfig}
\usepackage{amsmath,amsfonts,amsthm,amssymb}
\usepackage{physics}
\usepackage{float}
\usepackage[normalem]{ulem}

\usepackage[unicode]{hyperref}
\hypersetup{
 colorlinks,
 citecolor=blue,
 filecolor=blue,
 linkcolor=blue,
 urlcolor=blue
}

\begin{document}


\title{Effective aspect ratio of helices in shear flow}
\date{August 3, 2020}
\author{Brian Rost}
\author{Justin T. Stimatze}
\author{David A. Egolf}
\author{Jeffrey S. Urbach}
\email{urbachj@georgetown.edu}
\affiliation{
Department of Physics and Institute for Soft Matter Synthesis and Metrology,\\ Georgetown University, Washington, DC 20057.}

\begin{abstract}
We report the results of simulations of rigid colloidal helices suspended in a shear flow, using dissipative particle dynamics for a coarse-grained representation of the suspending fluid, as well as deterministic trajectories of non-Brownian helices calculated from the resistance tensor derived under the slender-body approximation. The shear flow produces nonuniform rotation of the helices, similarly to other high aspect ratio particles, such that more elongated helices spend more time aligned with the fluid velocity. We introduce a geometric effective aspect ratio calculated directly from the helix geometry and a dynamical effective aspect ratio derived from the trajectories of the particles and find that the two effective aspect ratios are approximately equal over the entire parameter range tested. We also describe observed transient deflections of the helical axis into the vorticity direction that can occur when the helix is rotating through the gradient direction and that depend on the rotation of the helix about its axis.
\end{abstract}

\maketitle
   
\section{\label{sec:intro} Introduction}
The behavior of flowing suspensions of fibers and other high aspect ratio particles plays an important role in a wide range of commercially important processes and consequently has been intensively investigated \cite{Lundell_Fluid_2011, Wang_Shear_2014}. Fibers with intrinsic curvature, including those with helical shapes, appear in many contexts \cite{Wang_Shear_2014,zhang1998biomimetic, kagawa1985fracture}, and their behavior under flow represents an interesting and challenging fundamental problem \cite{Leal_Particle_1980}.

A number of studies have specifically investigated the behavior of rigid helices in shear flow experimentally and computationally \cite{Kim_Separation_1991, Makino_Migration_2005, Kostur_Chiral_2006, Marcos_Separation_2009, Chen_Dynamical_2011, Zhang_Chirality_2017}. Like other high aspect ratio particles such as rods or ellipsoids, shear flow will cause helical filaments to rotate about an axis perpendicular to the velocity gradient and the flow directions (the vorticity axis) with a non-uniform rotation rate, with the particles spending more time with their long axis parallel to the flow than parallel to the shear gradient. For rotationally symmetric ellipsoids at low Reynolds number and in the absence of Brownian motion, the trajectories represent closed orbits (Jeffery Orbits) with analytic solutions that depend only on the particle's aspect ratio and its orientation with respect to the gradient axis, with no net motion in the vorticity direction \cite{dhont2007}. More generally, most axisymmetric bodies with fore-aft symmetry will also follow Jeffery Orbits in shear flow, with an effective aspect ratio that is determined by the square root of the ratio of the torque exerted on the body when it is held at rest with its axis along the gradient direction to the torque with its axis along the flow direction \cite{ISI:A1971I615100001}, although there are interesting exceptions in special cases \cite{Singh2013}.

Helical particles, by contrast, do not follow closed orbits, even in the absence of thermal fluctuations \cite{Kim_Separation_1991, Kostur_Chiral_2006}. Furthermore, the particles experience a net drift in the vorticity direction with a sign dependent on the helicity, a phenomenon which has been exploited to use shear flows to separate chiral objects \cite{Kim_Separation_1991, Kostur_Chiral_2006, Marcos_Separation_2009, Chen_Dynamical_2011, Kramel_Preferential_2016, Ro_Chiral_2016}. While considerable progress has been made understanding the average long time behavior of helices in shear flow in the presence of thermal fluctuations, we currently lack the ability to predict the short term dynamics of helical filaments, information that is critical for understanding the role of helical particles in suspension rheology or the behavior of particles in complex flows such as turbulence \cite{Kramel_Preferential_2016}.

In this work, we report the results of Dissipative Particle Dynamics (DPD) computer simulations of rigid helices in the presence of a shear flow. The DPD technique produces stochastic forces similar to thermal fluctuations and we observe that the orbits of the helices are qualitatively similar to noisy Jeffery Orbits. We compare the simulated trajectories with the deterministic trajectories calculated from the equation of motion for non-Brownian helices in the slender body approximation and report analytic expressions for the forces, torques, resistance tensor, center of mass velocity and angular velocity of a general helix in arbitrary orientation. We derive an analytic expression for a geometric {\em effective aspect ratio} calculated directly from the helix geometry and compare that to a dynamical effective aspect ratio calculated from the trajectories of the helices in the simulations and the deterministic trajectories. Over the entire parameter range tested, the geometric aspect ratio matches the measured dynamical aspect ratio in the simulations, within the statistical uncertainty, and accurately predicts the dynamical aspect ratio calculated from the deterministic trajectories. Finally, we discuss the origins of transient deflections of the helical axis into the vorticity direction that occur while the helix is rotating through the gradient direction in some, but not all, of the trajectories.
\section{\label{sec:methods} Methods}
\subsection{\label{sec:methods1} Simulations}

There have been many simulations of fibers in fluid flow using various approximations of hydrodynamic and contact interactions, employing a variety of techniques available with varying degrees of complexity and accuracy \cite{Bolintineanu:2014dj}. For these studies, we have used Dissipative Particle Dynamics (DPD) \citep{Hoogerbrugge1992, Groot1997}, an efficient coarse-grained fluid representation that can capture many aspects of the complex hydrodynamic interactions between the helical filament and the surrounding fluid and the effects of thermal fluctuations \citep{Fan2006, Duong-Hong2006} and is relatively simple to implement. The DPD implementation is similar to one we have used previously to study shear induced aggregation of straight rods \cite{Stimatze:2016jw} and is briefly summarized below.

In DPD, the coarse-grained fluid is represented by soft particles interacting via three pairwise forces: a repulsive force that determines the compressibility of the fluid, a dissipative force that models viscous dissipation, and a random force that determines the steady state temperature of the system. 

Thus the total force on particle $i$ is
\begin{eqnarray*}
{\bf F}_i = \sum_{j \neq i} \left(F_{ij}^{C} + F_{ij}^{R} + F_{ij}^{D} \right)\hat{\vb*{r}}_{ij},
\end{eqnarray*}
where
\begin{eqnarray*}
F^C_{ij}=Aw(r_{ij}),
\end{eqnarray*}
is the conservative, soft repulsion contribution to the force exerted by particle $j$ on particle $i$, where ${\vb*{r}}_{ij} = {\vb*{r}}_i - {\vb*{r}}_j$ 
and $r_{ij} = |{\vb*{r}}_{ij}|$.
The dissipative force is
\begin{eqnarray*}
F_{ij}^{D} = -\gamma w^2(r_{ij}) (\hat{\vb*{r}}_{ij} \cdot {\vb*{v}}_{ij}),
\end{eqnarray*}
where $\vb*{v}_{ij}=\dd {\vb*{r}}_{ij}/\dd t$, and the random force is
\begin{eqnarray*}
F_{ij}^{R} = \sigma w(r_{ij}) \alpha_{ij}/\sqrt{\Delta t},
\end{eqnarray*}
where $\alpha_{ij}$ is a random variable with unit variance Gaussian statistics, and $w(r)$ is a weighting function given by
\begin{eqnarray*}
w(r)= \left\{
\begin{array}{lr}
1 - r/r_c & r \leq r_c\\
0& r > r_c.
\end{array}
\right.
\end{eqnarray*}

Following Groot \& Warren \cite{Groot1997}, we set the repulsive parameter $A = 18.75$, the density $\rho = 4$, the random force coefficient $\sigma = 3$, the force cutoff radius $r_c=1$, and the dissipation coefficient $\gamma = 4.5$. The combination of $\rho$ and $A$ determines the compressibility, and values chosen are consistent with the compressibility of water. The combination of the strength of the random force $\sigma$ and the dissipation coefficient $\gamma$ determines the steady state kinetic energy (effective temperature) of the DPD system. For the values used here, $T=1$ in simulation units, which determined the appropriate timestep for the calculations (we use $\Delta t = 0.01$)\cite{Groot1997}. 
All numerical values are given in simulation units, with the relevant length scale being the particle size (unit diameter) and the time scale set by the applied shear. 

With parameters in this range, DPD has been shown to reproduce correct hydrodynamics at long length scales \cite{Espanol1995}. 
The helix length scales simulated in this work are not large compared to the DPD particle size, however, so quantitative agreement with Navier-Stokes hydrodynamics is not expected. 

Shear flow is generated by directly simulating moving boundaries at the top and bottom of the simulated fluid. 
\begin{figure}
	\centering
	\includegraphics[width=\columnwidth]{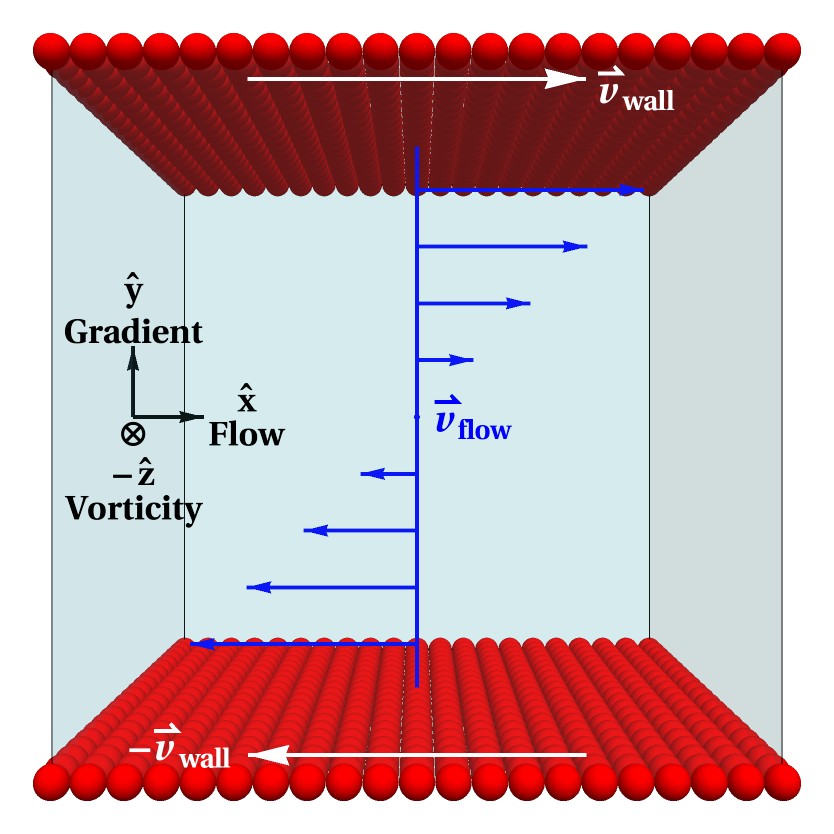}
	\caption{Schematic of simulation domain. Fixed walls one particle diameter thick in the horizontal $(x,z)$ plane (red spheres) are moved at constant, opposite speeds in the $x$ direction, thus producing simple shear with $(x, y, -z)$ = (flow, gradient, vorticity) directions. Periodic boundaries are employed in the $x$ and $z$ directions. }
	\label{simulation_setup}
\end{figure}
Specifically, referring to Fig.\ref{simulation_setup}, 
fixed walls one particle diameter thick in the horizontal $(x,z)$ plane are moved at constant, opposite speeds ($v_\text{wall}=1.5$) in the $x$ direction, thus producing simple shear with $(x, y, -z)$ = (flow, gradient, vorticity) directions. Periodic boundaries are employed in the $x$ and $z$ directions. For the simulations below, the simulated domain is cubic with a size that is adjusted according to the parameters of the helix being simulated to minimize self-interactions across periodic boundaries as well as contact with walls. The side length of the box was normally set as 1.5 times the helix length, rounded up to the nearest increment of 5 units.
No contact with walls was observed for any simulations reported here. 

\begin{figure}
	\centering
	\includegraphics[width=\columnwidth]{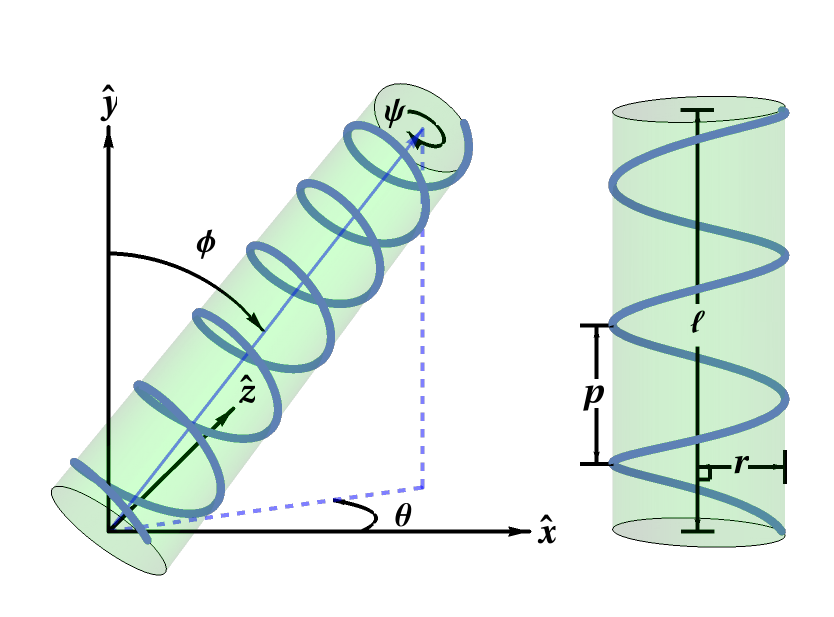}
	\caption{ Coordinates used for measuring helix orientation relative to shear flow (Fig. \ref{simulation_setup}) and parameters used for specifying helices.}
	\label{coord_param} 
\end{figure}

Helices are simulated by rigid strands of spherical particles with centers separated by a fixed spacing of half-unit length in simulation units. The particles interact with the fluid particles by the same DPD interactions described above. The initial helix configuration was set to be

\begin{equation}\label{eq:helix}
\begin{pmatrix}
x(u)\\
y(u)\\
z(u)
\end{pmatrix}=
\begin{pmatrix}
r \cos(2\pi nu/\ell_c)\\
\ell u/\ell_c\\
h r \sin(2\pi nu/\ell_c)
\end{pmatrix}
\!\!\!\qq{with}\!\!\! \begin{array}{l}
0\leq u\leq \ell_c\\h=\pm1,
\end{array}
\end{equation}

which gives a left(right)-handed helix when $h=1$($h=-1$) of filament length $\ell_c=\sqrt{\ell^2+4 \pi^2 n^2 r^2}$ with $n$ turns, radius $r$, end-to-end length $l$ and pitch $p=l/r$, initially oriented parallel to the gradient direction such that a perpendicular from the helical axis to the first bead on the $+y$ end of the helix was in the positive flow direction ($(\phi,\theta,\psi)=(0,0,0)$, see Fig. \ref{coord_param}). All parameter sets were run for 50,000 timesteps, then extended until at least 2 ``flips'' were observed, representing at least one full orbit. 

One hundred and five simulations of isolated helices with varying pitch ($p$), radius ($r$) and length ($\ell$) were performed using the Large-scale Atomic/Molecular Massively Parallel Simulator (LAMMPS) \cite{Plimpton1995} environment, including the LAMMPS implementation of DPD and the LAMMPS rigid-body integrator for calculating net forces and torques on the helices.
The resulting dynamical equations were solved using the velocity-Verlet integration scheme.

\subsection{\label{sec:methods2} Analytic Methods}

We also present results from deterministic trajectories calculated with the slender body approximation, appropriate for thin filaments in Stokes (low Reynolds number) flow. The equations of motion are determined from the resistance tensor for the helix 
under the conditions of zero net force and torque.

The force on the helix from slender body theory is given by
\begin{equation}\label{eq:f0}
\vb{F} = \int_{0}^{\ell_c}\dd{u}\left(\alpha_t \vb*{v}_r^t+\alpha_n \vb*{v}_r^n\right),
\end{equation}
and the torque by
\begin{equation}\label{eq:t0}
\vb*{\tau} = \int_{0}^{\ell_c}\dd{u}\vb*{r}\cross\left(\alpha_t \vb*{v}_r^t+\alpha_n \vb*{v}_r^n\right),
\end{equation}
where $\vb*{v}_r$ is the velocity of the fluid relative to the helical fiber, with superscript $t \ (n)$ representing the component tangent (normal) to the fiber, $\alpha_t \ (\alpha_n)$ is the drag coefficient for flow tangent (normal) to the fiber, and $\vb*{r}$ is the vector from the helix center to a point on the fiber. Being in a Stokes flow regime immediately implies that $\vb{F}=\vb*{\tau}=\vb{0}$.

We can break $\vb*{v}_r$ into contributions from the flow and from the helix's motion as $\vb*{v}_r = \vb*{v}_f-\vb*{v}_{cm}-\vb*{\omega}\cross\vb*{r}$ where $\vb*{v}_f$ is the fluid velocity, $\vb*{v}_{cm}$ is the helix center of mass velocity, and $\vb*{\omega}$ is the angular velocity of the helix. This allows us to recast Eq. \ref{eq:f0} and \ref{eq:t0}, defining $\vb*{v}_h\equiv\vb*{v}_{cm}+\vb*{\omega}\cross\vb*{r}$ and dropping the integration bounds, as

\begin{equation}\label{eq:f1}
\int\dd{u}\left(\alpha_t \vb*{v}_f^t+\alpha_n \vb*{v}_f^n\right)=\int\dd{u}\left(\alpha_t \vb*{v}_h^t+\alpha_n \vb*{v}_h^n\right),
\end{equation}
and
\begin{equation}\label{eq:t1}
\int\dd{u}\vb*{r}\!\cross\!\left(\alpha_t \vb*{v}_f^t+\alpha_n \vb*{v}_f^n\right)=\!\int\dd{u}\vb*{r}\!\cross\!\left(\alpha_t \vb*{v}_h^t+\alpha_n \vb*{v}_h^n\right).
\end{equation}
Together these two equations may be written as a single matrix equation
\begin{equation}\label{eq:mat}
\begin{pmatrix}
\vb{F}_f \\ \vb*{\tau}_f
\end{pmatrix}
=
\overline{\overline{\vb{R}}}
\cdot
\begin{pmatrix}
\vb*{v}_{cm} \\ \vb*{\omega}
\end{pmatrix},
\end{equation}
where $\vb{F}_f$ and $\vb*{\tau}_f$ are the force and torque on a motionless (i.e. anchored in place) helix due to the unperturbed fluid flow, and $\overline{\overline{\vb{R}}}$ is the $6\times6$ symmetric resistance tensor for the helix. We are able to analytically find $\vb{F}_f$, $\vb*{\tau}_f$, and $\overline{\overline{\vb{R}}}$, then solve Eq. \ref{eq:mat} to find $\vb*{v}_{cm}$ and $\vb*{\omega}$ for an arbitrary helix in an arbitrary orientation. 

A helix in an arbitrary orientation can be described by the vector $\vb{H}$, obtained by applying rotation matrices for the three angles $(\phi,\theta,\psi)$ to the vector described by Eq. \ref{eq:helix}.

The drag on the segments of the filament that compose the helix is anisotropic. From slender-body theory at low Reynolds number, the drag coefficient for flow normal to the filament is twice the drag coefficient for flow tangent to the filament, i.e. $\alpha_n/\alpha_t=2$ \cite{batchelor_1970}. Allowing for a nonzero filament thickness results in $\alpha_n/\alpha_t=2$ to within a logarithmic correction term \cite{Marcos,Childress}. We take $\alpha_n=2\alpha_t$ following slender body theory throughout the rest of the paper unless explicitly stated otherwise.

We decompose the unperturbed fluid flow, given by $\vb*{v}_f=\dot{\gamma}(\vb{H}\cdot \hat{y})\hat{x}$, into tangent, $\vb*{v}_f^t=(\vb*{v}_f\cdot\hat{t})\hat{t}$, and normal, $\vb*{v}_f^n=\vb*{v}_f-\vb*{v}_f^t$, components 
where $\hat{t}$ is a unit vector parallel to the filament (i.e. in the direction of $\dv{\vb{H}}{u}$). The drag force per unit length is given by
\begin{equation}
\vb*{f}_d = \alpha_t\vb*{v}_f^t+\alpha_n \vb*{v}_f^n=\alpha_n \vb*{v}_f+\left(\alpha_t-\alpha_n\right)\vb*{v}_f^t,
\end{equation}
and so the force and torque on the fixed helix are
\begin{equation}
\vb{F}_f=\int_{0}^{\ell_c}\dd{u}\vb*{f}_d\qq{and}\vb*{\tau}_f=\int_{0}^{\ell_c}\dd{u}\vb{H}\cross\vb*{f}_d.
\end{equation}
The resulting expressions are very messy, but can be easily evaluated for any helix parameters and orientation, and are provided in the Supplemental Material \cite{supplemental}.

Using the same method as described above, we decompose the helix velocity, $\vb*{v}=\vb*{v}_{cm}+\vb{H}\cross\vb*{\omega}$, into normal and tangent components. This allows us to evaluate the RHS of Eqs. \ref{eq:f1} and \ref{eq:t1} from which $\overline{\overline{\vb{R}}}$ may be directly read off from the coefficients of $\vb*{v}_{cm}$ and $\vb*{\omega}$. The full expression for $\overline{\overline{\vb{R}}}$ is provided in the Supplemental Material \cite{supplemental}. To our knowledge, this is the first explicit expression for the resistance tensor of an arbitrary helix in an arbitrary orientation.

We then solve Eq. \ref{eq:mat} for $\vb*{v}_{cm}$ and $\vb*{\omega}$. The resulting expressions are too cumbersome to report but are available in the Supplemental Material as a Mathematica binary \cite{supplemental}. Having analytic expressions for the center of mass and angular velocities enabled us to numerically integrate the coupled differential equations governing the dynamics of the helix at a significantly lower computational cost than was previously possible. We are thus able to compare the noisy DPD simulations with the dynamics determined from numerically integrating the solution to the deterministic equations of motion for Stokes flow in the slender body approximation.

\section{\label{sec:torque} Geometric effective aspect ratio}
As described in the introduction, the trajectories of an axisymmetric body with fore-aft symmetry in viscous shear flow are given by Jeffery Orbits that are fully determined by the ratio of the torque exerted on the body when it is held at rest with its axis along the gradient direction to the torque with its axis along the flow direction \cite{ISI:A1971I615100001}. Inspired by this result, we use this quantity for helices to define a geometric effective aspect ratio for helices, despite the fact that they are {\em not} axisymmetric, nor do they have fore-aft symmetry.

\begin{figure}
	\centering
	\includegraphics[width=\columnwidth]{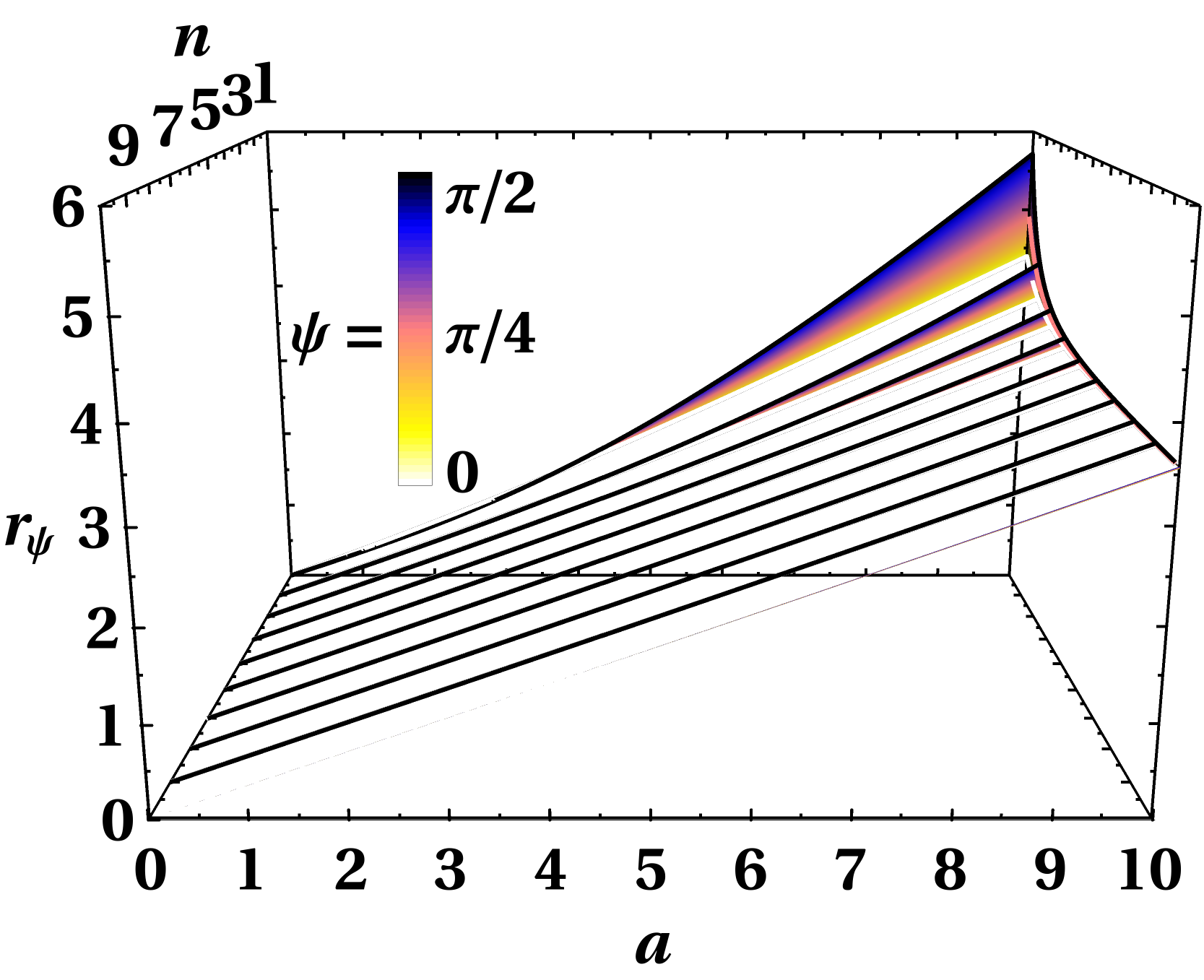}
	\caption{Geometric effective aspect ratio ($r_{\psi}$ from Eq. \ref{rGpsi}) as a function of $a=\ell/r$ for integer $n$ from $n=1$ to $n=10$. The value of $\psi$ is indicated by color. The dependence on $\psi$ is only significant at small $n$ (large pitch) and large $a$.}
	\label{r_Gint}
\end{figure}

Following Cox \cite{ISI:A1971I615100001}, we define a ($\psi$-dependent) geometrical aspect ratio for the helix as 
\begin{equation}
r_{\psi}=\sqrt{\frac{\tau_1}{\tau_2}}=\sqrt{\frac{2}{3}} \frac{\ell}{r} \sqrt{\frac{3 r^2(\pi ^2 n^2 +\cos (2 \psi ))+\ell ^2}{(4 \pi n r)^2+\ell ^2(2+\cos (2 \psi ))}},
\label{rGpsi}
\end{equation}
where $\tau_1$ is $\vb*{\tau}_f\cdot\hat{z}$ (as defined in Eq. \ref{eq:t0}) when the helical axis is parallel to the gradient direction $\hat{y}$ ($\phi=0, \ \theta=0$) and $\tau_2$ is $\vb*{\tau}_f\cdot\hat{z}$ when the helical axis is parallel to the flow direction $\hat{x}$ ($\phi=\pi/2, \ \theta=0$).\\\\
Unlike the axisymmetric bodies studied by Cox, this definition does not uniquely specify the aspect ratio for a given geometry due to the dependence on the angle $\psi$. Moreover, given that $\psi$ will change during an orbit, the situation is clearly more complicated, and the simple ratio of torques at fixed position will not be sufficient to precisely determine the trajectory. Nonetheless, we can still use Eq. \ref{rGpsi} as an aspect ratio that depends only on the geometry and $\psi$ and investigate the extent to which that quantity is a useful predictor of the actual trajectories.  Figure \ref{r_Gint} shows how $r_{\psi}$ (Eq. \ref{rGpsi}) varies with $a=\ell/r$, $n$ and $\psi$. 

Our goal is to find an aspect ratio based solely on the geometry of the helix 
that is a good predictor of the fraction of time helices spend aligned with the shear flow for a wide range of helical parameters. 
We note that for large $n$ and $a$, $r_{\psi}\approx {a}/({2\sqrt{2}}) \approx 0.35 a$, with significant deviations only for small $n$ (large pitch). Indeed $\lim_{n\to\infty}r_\psi=a/(2\sqrt{2})$ for all $a$ and $\psi$. This suggest an approximate geometrical aspect ratio based solely on $a$, $r_a=a/(2\sqrt{2})$. As can be seen in Fig. \ref{r_Gint}, $r_\psi \approx r_a$ for $n \geq 5$. However, as discussed below, we find that $r_a$ is not a good predictor of the fraction of time helices spend aligned with the shear flow for much of the parameter range consider here. 

It is perhaps not surprising that helix parameters beyond the bounding cylinder aspect ratio need to be taken into account to accurately describe the tumbling trajectories.    Thus we introduce an alternative purely geometrical aspect ratio, which we denote as $r_G$, produced by taking $\psi=\pi/4$ in Eq. \ref{rGpsi}. 
$r_G$ can be calculated from Eq. \ref{rGpsi} and  is given by

\begin{equation}
r_G=r_{\psi=\pi/4}=\frac{a}{\sqrt{3}} \sqrt{\frac{a^2+3 \pi ^2 n^2}{a^2+8 \pi ^2 n^2}}.
\label{rG}
\end{equation}

Below we show that $r_G$  works quite well as a predictor of the fraction of time helices spend aligned with the shear flow for all of the parameters studied here, despite the variation of $\psi$ during a trajectory. 


\section{\label{sec:results} Results}

\subsection{\label{sec:Jeffery} Jeffery-like Orbits }
\begin{figure}
	\centering
\includegraphics[width=\columnwidth]{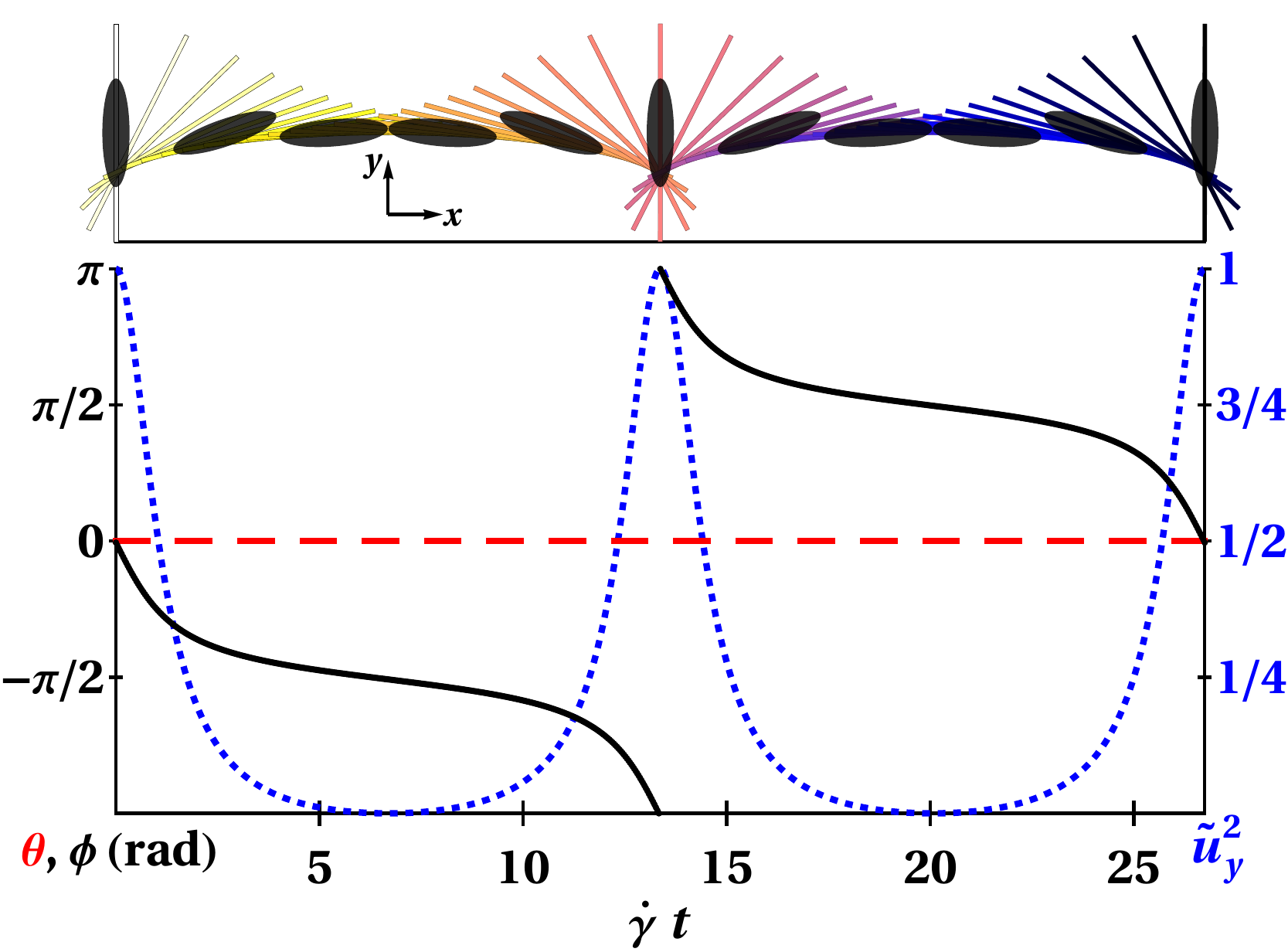}	
\caption{Jeffery Orbit for a non-Brownian ellipsoid in a shear flow. The top image shows successive orientation snapshots ($\tilde{u}$) for a particle of aspect ratio 4, with a small center of mass velocity from left to right for clarity, colored from white (early times) through black (late times). The graph displays angles relative to gradient ($\phi$, solid black) and vorticity ($\theta$, dashed red) directions and square of the gradient ($y$) component of the orientation unit vector ($\tilde{u}_y^2$, dotted blue) vs. strain. 	}
	\label{jorbits}
\end{figure}

Trajectories for isolated helices were initialized with their helical axis oriented in the gradient direction. In this configuration the shear flow exerts a torque on the helix parallel to the vorticity axis, resulting in a rapid rotation into the flow direction. The torque is reduced as the helix aligns with the flow, so the rotation rate decreases, reaching a minimum when the axis of the helix is perpendicular to the gradient ($y$) axis. This behavior is qualitatively similar to Jeffery Orbits of non-Brownian axisymmetric ellipsoids, which are deterministic, closed orbits with a fixed angle, $\theta$, relative to the vorticity direction. As an example, Fig. \ref{jorbits} displays an orbit for an ellipsoid of revolution with aspect ratio $r_e=a/2=4$ in a series of orientation snapshots and plots of $\phi$ and $\theta$ vs. time.

An alternative way to visualize the trajectories is to track the evolution of components of the orientation vector, $\hat u$, a unit vector aligned with the axis of the helix. We define $\tilde{u}$, the normalized projection of $\hat{u}$ onto the flow-gradient ($xy$) plane, to more easily visualize the rotation of the helix about the vorticity ($-z$) axis. It is then easy to characterize the orbiting behavior of the helix using $\tilde{u}_y\equiv\tilde{u}\cdot\hat{y}= \hat{u}\cdot\hat{y}/\cos{\theta}=\cos{\phi}$ and the deflections into the vorticity direction by
$u_z\equiv\hat{u}\cdot\hat{z}=\sin{\theta}$. The dotted blue curve in Fig. \ref{jorbits} shows $\tilde{u}_y^2$, vs. strain ($\dot\gamma t)$, where the relatively slow rotation rate when the particle is aligned in the flow direction produces an extended period of time when $u_y^2$ is small. 

\begin{figure}
	\centering
\includegraphics[width=\columnwidth]{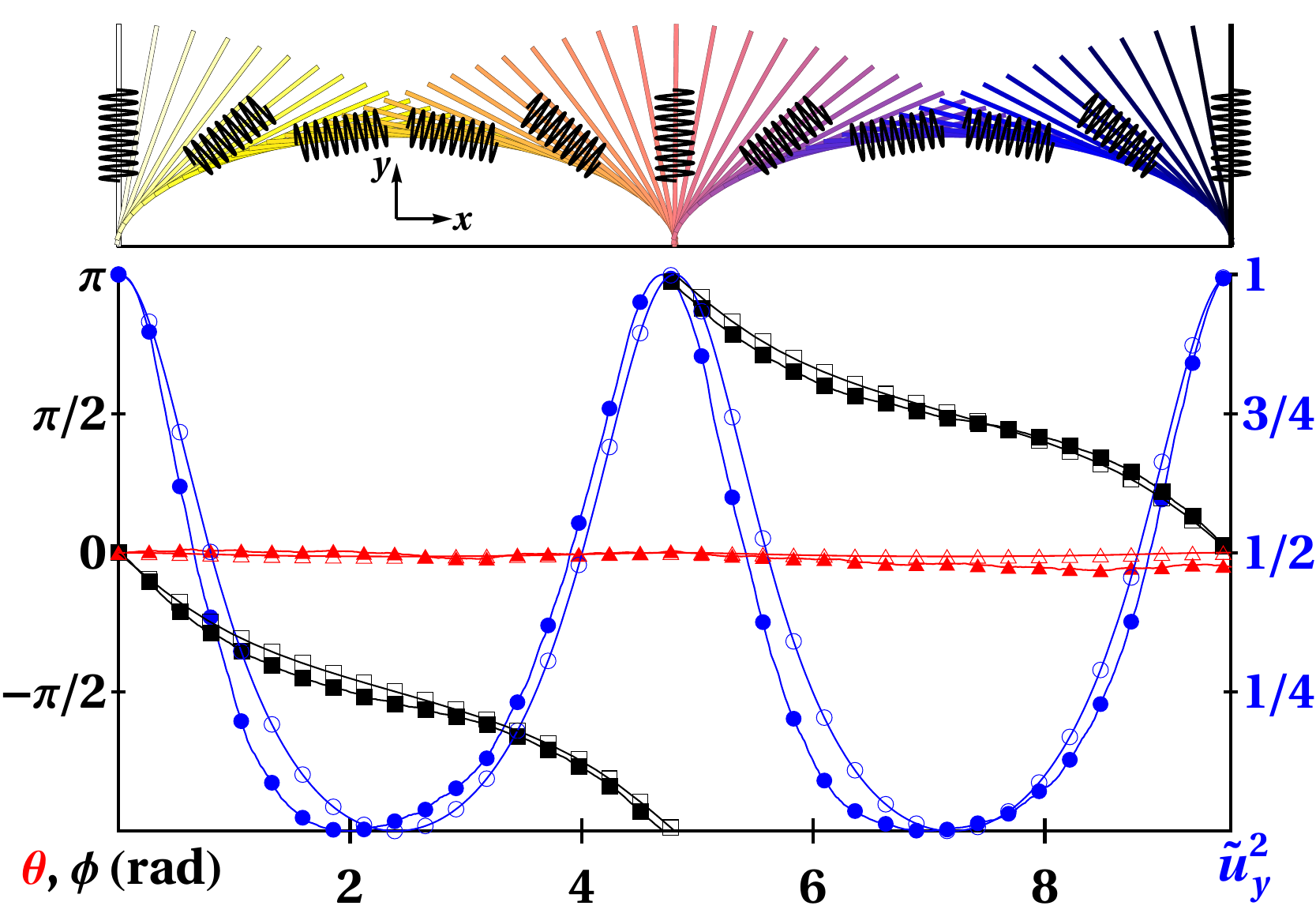}
\includegraphics[width=\columnwidth]{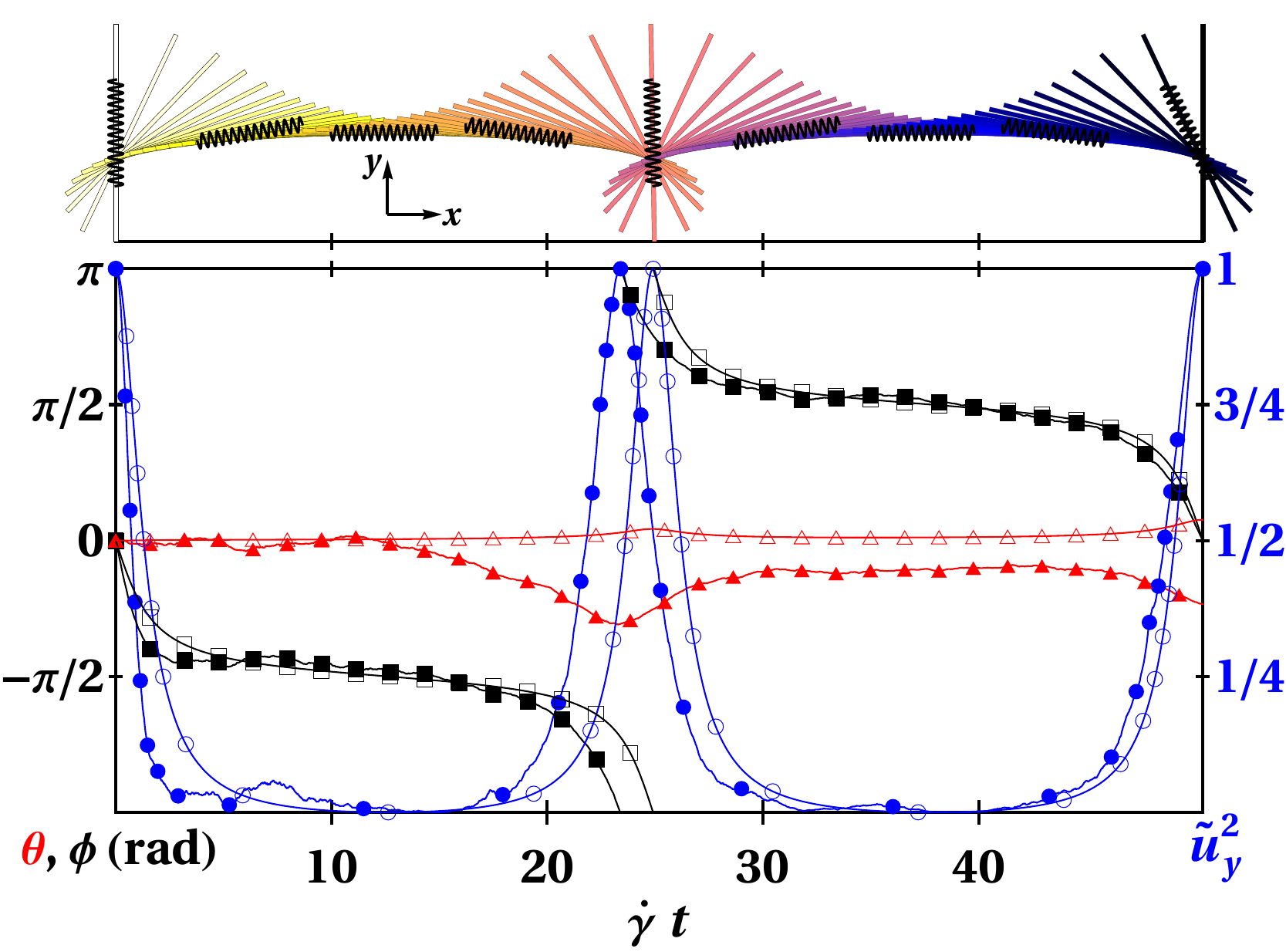}
\caption{Trajectories of helical filaments from DPD simulations and deterministic calculations. Top: Snapshots of $\tilde{u}$ 
as in Fig. \ref{jorbits} from deterministic trajectory of a helix with $a=5$ and $n=9$, and $\phi$, $\theta$, and $\tilde{u}_y^2$, from the deterministic (open) and simulated (filled) trajectories. Bottom: Orientation snapshots, $\phi$, $\theta$, and $\tilde{u}_y^2$ for a helix with $a=15$, $n=14$
	}
	\label{trajectories}
\end{figure}

Figure \ref{trajectories} displays results from two representative helical geometries, each showing a deterministic calculation (open circles) and a DPD simulation (filled), revealing both the non-uniform rotation rates and the stochastic variations that arise in the DPD simulations.
As expected, we observe that squat helices, with smaller length to radius ($a=\ell/r$) values, display relatively small variation in their rotation rates (top panel, $a=5$), whereas helices with high values of $a$ show rotation rates that slow down dramatically when aligned in the flow direction (bottom panel, $a=15$). 

For the parameter regimes investigated here, the stochastic component of the motion in the simulations is relatively small, suggesting that the effective rotational Peclet number, the ratio of the shear rate to the rotation diffusion coefficient, $D_r$, is large. We reported previously that for a rigid rod of 21 particles under identical conditions, $D_r = 4 \times 10^{-6}$ in simulation units \cite{Stimatze:2016jw}, and in general for rods, $D_r \propto L^{-3}$. 

\subsection{\label{sec:AR} Effective Aspect Ratio }

For ideal Jeffery Orbits, the angle $\phi$ of the particle relative to velocity direction is related to the aspect ratio $r_e$ according to 
\begin{equation}
r_e = \frac{1}{\langle \cos^2\phi \rangle} - 1,
\label{eq:r_e}
\end{equation}
where brackets represent the long time average. Experimentally or computationally accessing the long time average is clearly difficult, but we observe that from Jeffery's equations (and as visualized in Fig. \ref{jorbits}), $\cos^2\phi$ is periodic with a period of half an orbit. Furthermore, each half orbit is mirror symmetric about its midpoint. This means that the long time average, for particles obeying Jeffery's equations, of $\cos^2\phi$ is equal to the average over an integral number of quarter orbits.

A spherical particle $(r_e=1)$ has a uniform rotation rate, producing $\langle \cos^2\phi \rangle =1/2$. As $r_e$ increases, the particle spends a longer fraction of its orbit aligned in the flow direction, so $\langle \cos^2\phi \rangle$ (and $\langle \tilde{u}_y^2 \rangle$) decreases. Note that this result is independent of $\theta$, the angle with respect to the gradient-velocity plane (which is constant for a Jeffery Orbit and therefore determined uniquely by the initial conditions).


Equation \ref{eq:r_e} can be easily generalized to calculate an effective aspect ratio from our simulated trajectories (as in \cite{Stimatze:2016jw}), with a couple of caveats. One is that $\theta$ is not constant, varying due to both thermal fluctuations and torques with components in the flow-gradient plane. When $\theta$ approaches $\pi/2$ (helix aligned in the vorticity direction), thermal noise causes $\phi$ to fluctuate erratically. This issue does not create difficulties in this work, where we focus on relatively short trajectories with initial conditions of $\theta=0$. A second caveat is that, as discussed above, the effective aspect ratio calculation from measured trajectories requires an integral number of quarter orbits,
which can create selection bias for finite length trajectories. 
In order to minimize this effect, we have included only parameter ranges where all simulated trajectories included at least one full rotation. We further discard the first half rotation to mitigate any possible transient effects in the simulation.

\begin{figure*}[ht!]
\centering
 \includegraphics[width=\textwidth]{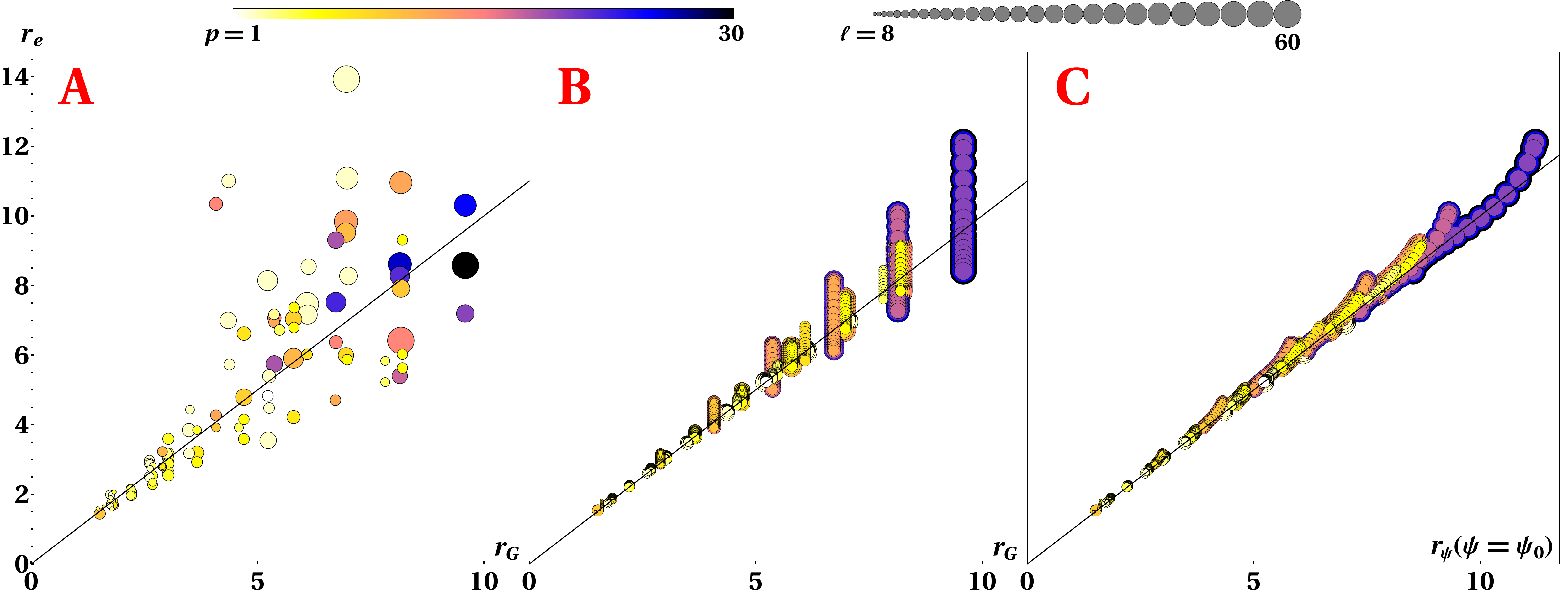}
 \caption{A) Plot of the effective aspect ratio $r_e$ (Eq. \ref{eq:r_e}) from simulations versus the geometric aspect ratio $r_G$ (Eq. \ref{rG}) for a range of helix parameters. Marker size is indicative of helix length, (min,max) = (8,60) and marker color is indicative of pitch with white corresponding to $p=1$ through black (dark) with $p=30$.  The solid line is $r_e=r_G$ (all 3 panels). B) Plot of $r_e$ from deterministic trajectories versus $r_G$ for a range of helix parameters and different initial values of $\psi$ C) Data from (B) plotted vs. $r_{\psi}$ (Eq. \ref{rGpsi}), calculated using the initial values of $\psi$ for that run.}
 \label{AR}
\end{figure*}

Here we seek to determine if the empirical effective aspect ratio $r_e$, calculated from Eq. \ref{eq:r_e}, can be simply related to geometric parameters of the helix. Figure \ref{AR}A shows a scatter plot of $r_e$, determined from the simulations, versus $r_G$, defined by Eq. \ref{rG}, for a range of helix parameters. Although there is considerable scatter in the data, there is a strong correlation between the two quantities, with no evident dependence on either $\ell$ or $p$ (indicated by the size and color of the points, respectively). 
A least squares fit to $\log{r_e}$ vs. $\log{r_G}$ produces $r_e=1.01r_G^{1.02}$ and is consistent with the two quantities being equivalent ($r_e=Ar_G^\alpha$, with $\alpha =\{0.93,1.10\}$ and $A=\{0.89,1.15\}$ at a 95\% confidence level (Cl).

While the geometric aspect ratio defined by Eq. \ref{rG} does a remarkably good job of predicting $r_e$, it is worth noting that, as discussed above, over much of the parameter range, $r_G$ is close to $r_a={a}/({2\sqrt{2}})$ (see Fig. \ref{r_Gint}). A plot of $r_e$ vs $r_a$ looks qualitatively similar to Fig. \ref{AR}A, but shows systematic deviations for small $n$. Our results are inconsistent with $r_e=r_a$ ($r_e=A_1r_a^{\alpha_1}$, with $\alpha_1 = \{1.04, 1.22\}$ and $A_1=\{0.84, 1.08\}$ at a 95\% Cl.

We find that $r_e$ and $r_G$ are similarly closely related for the deterministic helix trajectories, as shown in Fig. \ref{AR}B, for the same helix parameters used in Fig. \ref{AR}A, but with initial values of $\psi$ distributed between 0 and $2\pi$. A least squares fit to $\log{r_e}$ vs. $\log{r_G}$ here produces $r_e=1.01 r_G^{1.01}$. The scatter in the data clearly shows that $r_e$ depends on $\psi$, suggesting that the scatter in Fig. \ref{AR}A arises from a combination of the effects of Brownian motion and the limitations of the definition of a $\psi$-independent $r_G$. Fig. \ref{AR}C shows the result of plotting the same data for $r_e$ versus $r_{\psi}$, the $\psi$-dependent geometric aspect ratio calculated from Eq. \ref{rGpsi}, using the initial value of $\psi$ for each trajectory. This appears to reduce the scatter and gives a least squares fit to $\log{r_e}$ vs. $\log{r_\psi}$ as $r_e=1.01 r_\psi^{1.00}$. Note that $\psi$ can change significantly over the course of the orbit, as shown below.

These results show that $r_G$ provides a reasonably robust estimate of $r_e$ for both the DPD simulations and the deterministic trajectories calculated using slender body theory, but that the absence of axisymmetry manifests itself in a dependence on the angle $\psi$ that cannot be easily captured. To our knowledge, this is the first test of an effective aspect ratio of helices in shear flow. Marcos et al. \cite{Marcos_Separation_2009} numerically calculate effective aspect ratios for helical bacteria based on the ratio of rotation rates for $\phi=0$ and $\pi/2$, which produces very similar results to the $r_G$ defined here, but is a significantly messier expression \cite{supplemental}.

\subsection{\label{sec:vorticity} Deflections in Vorticity Direction }
\begin{figure}
\centering
 \includegraphics[width=\columnwidth]{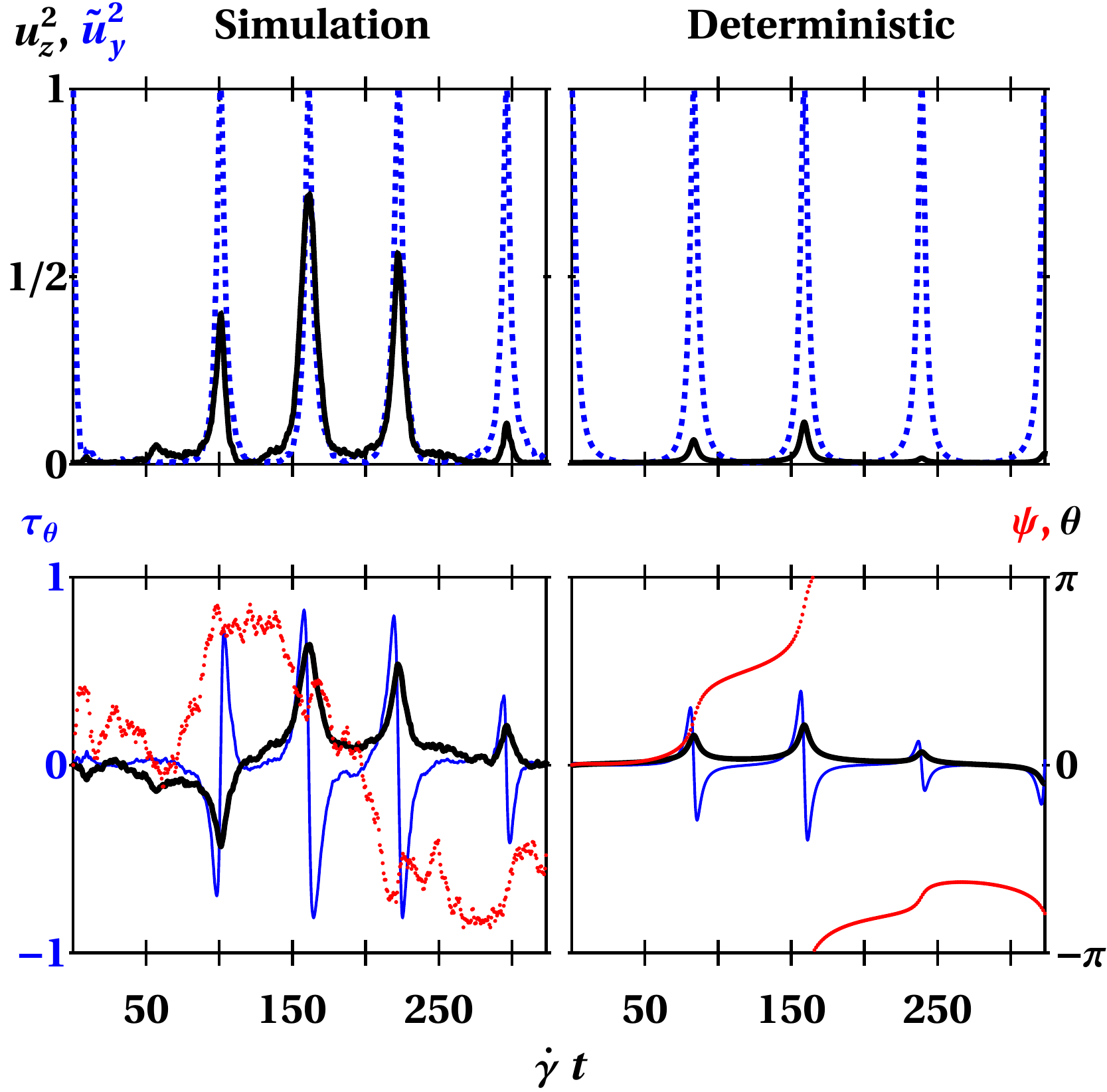}
 \caption{Top: plot of $\tilde u_y^2$ (dashed) and $u_z^2$ (solid) vs. time for one simulated (left) and one deterministic (right) trajectory ($\ell = 58.8$, $n=4$, $r=3$ for both) demonstrating deflection into the vorticity direction, and subsequent recovery, as the helix axis rotates past the gradient direction. Bottom: plot of $\psi(t)$ (dotted), $\tau_\theta$ in units of $\dot\gamma\alpha_n \ell_c^2 r$ (thin) and $\theta(t)$ (thick) for the same trajectories. The deflections into the vorticity direction are well predicted by $\tau_\theta$, while the variability in size and direction of deflection arises from the behavior of $\psi(t)$.}
 \label{vorticity}
\end{figure}

Although the primary focus of this study is the rotation about the vorticity axis, we close by noting an interesting behavior that occurs in many, but not all, of the simulated and deterministic trajectories. Figure \ref{vorticity} shows the evolution of $\tilde{u}_y^2$ and $u_z^2$ for a representative trajectory from the simulation, where we find a transient deflection in the vorticity direction that peaks when $\tilde{u}_y^2$ is maximum, i.e when $\phi=n\pi$ and the rotation rate of the Jeffery-like Orbit is at its maximum. Similar, albeit smaller, deflections are observed in the deterministic trajectory for the same initial condition. For both simulated and deterministic trajectories, the deflections vary in magnitude and sign, with no obvious pattern, and are sometimes absent altogether.

We can gain some insight into the origin of these transient deflections by considering the torque on a fixed helix (LHS of Eq. \ref{eq:t1}). The torque causing deflections into the vorticity direction is given by $\tau_\theta=\vb*{\tau}\cdot\hat{\theta}\equiv \vb*{\tau}\cdot\left(\hat{x} \cos (\phi )+\hat{y} \sin (\phi )\right)$. This quantity ($\tau_\theta$) is plotted in the lower panels of Fig. \ref{vorticity}, using the instantaneous values of $\phi$, $\theta$ and $\psi$, alongside plots of $\theta (t)$ and $\psi (t)$. It is clear that $\tau_{\theta}$ is highly correlated with $\dot\theta$, although it should be noted that the exact behavior of the vorticity deflections is characterized by $\omega_\theta\equiv\vb*{\omega}\cdot\hat{\theta}$ which is an \textit{extremely} complicated expression.

The full expression for $\tau_\theta$ itself is quite unruly, but we can capture some aspects of the relevant behavior if we restrict our attention to a helix in the gradient-velocity plane ($\tau_\theta$ at $\theta = 0$). This  quantity, which we label $\tau_{\theta}^0$, is found to be

\begin{widetext}
\begin{equation}\label{tautheta}
\tau_{\theta}^0=\frac{\dot\gamma\alpha _t r \ell \cos (\psi ) \left(h \sin (2 \phi ) \left(9 \ell ^2-6 \ell _c^2-4 \pi ^2 n^2 r^2 \cos ^2(\psi )\right)+3 \pi n r \ell \sin (\psi ) (\cos (2 \phi )+3)\right)}{12 \pi n \ell _c}.
\end{equation}
\end{widetext}

While the full expression for $\tau_{\theta}^0$ is still complex, in the limit of small $r$ it reduces to simply $\tau_{\theta}^0\approx \dot{\gamma}\alpha_t h \ell^2 r\cos (\psi ) \sin (2 \phi )/(4 \pi n)$. 
The $\sin(2 \phi)$ dependence produces a contribution that has extrema at $\phi=\pi/4$ and $3\pi/4$, with a sign change in between, and thus explains the transient deflection. The origin of this dependence can be understood by recalling that simple shear can be represented as a superposition of pure shear (or extensional flow) and pure rotational flow. The contribution from the rotational flow is $\phi$-independent (the very last term in Eq. \ref{tautheta}, of order $r^2$). The extensional flow is outward along the extensional axis ($\phi=\pi/4$) and inward along the compressional axis ($\phi=3\pi/4$), and thus is responsible for the $\sin(2 \phi)$ term in $\tau_{\theta}^0$. 
At $\phi=\pi/4$ and $3\pi/4$, the axial flow produces drag forces that are primarily parallel to the helical axis (although tilted towards the filament normal because of the anisotropic drag), but will contribute to a $\psi$-dependent $\tau_\theta^0$ because the finite radius of the filament means that each filament element generally has an $r_z$ that is non-zero, so $\vec{r}\cross \vec{F}$ can contain elements that contribute to $\tau_\theta^0$. Interestingly, we find that the expression for $\tau_{\theta}^0$ in the limit of small $r$ given above is also valid for isotropic drag ($\alpha _t =\alpha _n$), indicating that the general behavior can be understood from this geometric argument, without consideration of the tilting of the drag force relative to the flow direction. 

\begin{figure}
	\centering
\includegraphics[width=\columnwidth]{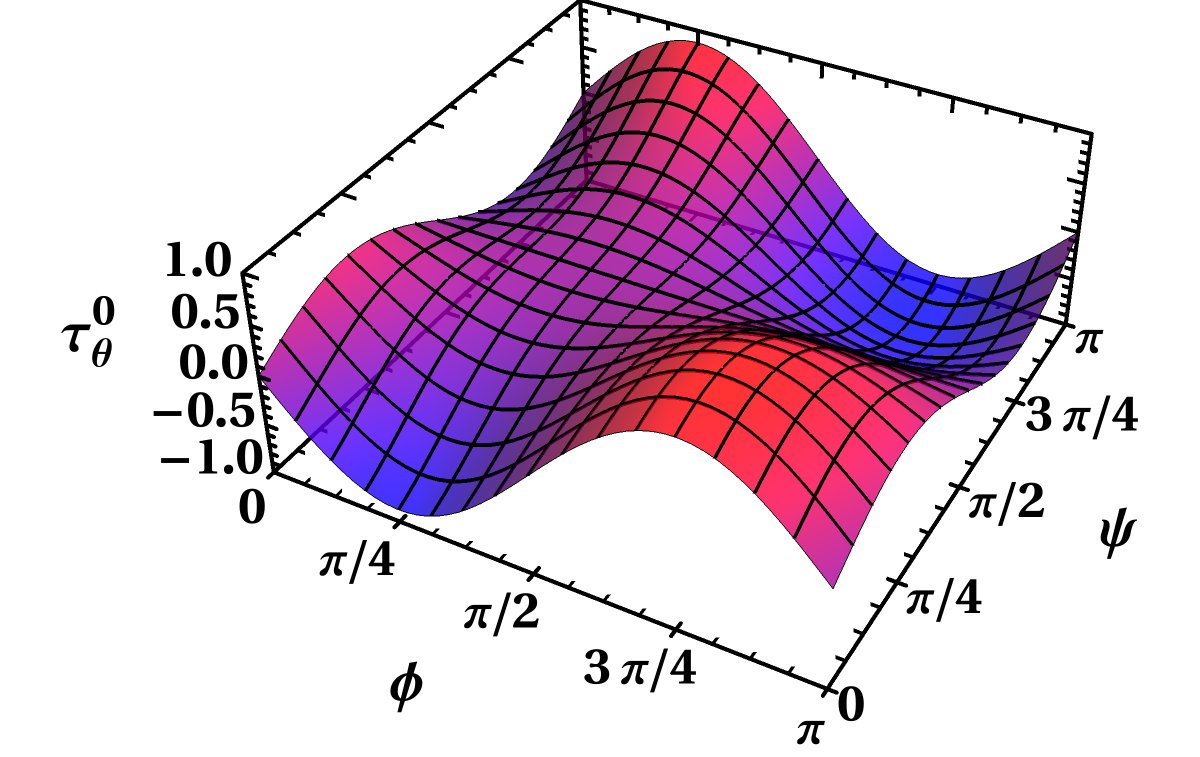}
\caption{Surface plot showing $\tau_\theta^0$ (in units of $\dot\gamma\alpha_n \ell_c^2 r$) as defined in Eq. \ref{tautheta} with $\ell=58.8$, $r=3$ and $n=4$ as a function of $\phi$ and $\psi$.}
	\label{vorticity_deflection}
\end{figure}

Fig. \ref{vorticity_deflection}, which shows a surface plot of $\tau_{\theta}^0(\phi,\psi)$ for a representative set of helix parameters, has a dominant $\sin(2 \phi)$ dependence that is modulated by a sinusoidal function of $\psi$. This likely underlies the variability in the sign and magnitude of the observed vorticity deflections. Furthermore, $\psi$ evolves continuously during the orbit, as can be seen in Fig. \ref{vorticity}, and this evolution presumably leads to the complexity of the helical trajectories even in the absence of Brownian motion. 

\section{\label{sec:discussion} Discussion}
As described in the Introduction, rigid helical filaments initially oriented parallel to a shear gradient will rotate about the vorticity axis with a rotation rate that decreases as the helix aligns in the flow direction. Our simulations and deterministic trajectories show, as expected, that the reduction of the rotation rate increases with the aspect ratio of the bounding cylinder ($a/2=l/2r$, Fig. \ref{coord_param}). We find that $a$ is the primary determinant of the degree of alignment and therefore that it is relatively insensitive to the other dimensionless ratios that describe a particular helix (such as the number of turns, $n=\ell/p$, and the pitch angle (related to $r/p$)), at least in the range simulated here ($4<a<20$, $1<n<29$, $0.07 < r/p < 2$).
(The thickness of helical filament itself,  one particle diameter, was not varied.) 

Qualitatively, this behavior can be understood as follows: when the helix is aligned in the gradient direction, the effect of the fluid drag will cause rotation about the vorticity axis with a rate that is comparable to the shear rate. When the helix is aligned in the velocity direction, the torque due to the shear flow, and therefore the rotation rate, is reduced, as with other high aspect ratio particles. The ratio of the torques in those two orientations is mostly determined by $a=\ell/r$. Decreasing the pitch $p$ (or, equivalently, increasing $n$) for a given $\ell$ will increase the torque overall, but not the asymmetry. 

We do find, however, that there are significant deviations from the degree of alignment that would be predicted by only considering $\ell/r$, particularly for small values of $n$. The geometric aspect ratio defined by Eq. \ref{rG}, based on the ratio of the torque exerted on the helix held at rest with its axis along the shear to the torque with its axis in the flow direction using slender body theory appropriate for low Reynolds number flow, accurately accounts for those deviations. In fact the data displayed in Fig. \ref{AR}A shows that $r_G$ is approximately equal to $r_e$, to within statistical uncertainty, over the entire parameter range simulated. However, there is considerable scatter in the data for large aspect ratio helices because of the relatively small number of complete orbits observed, leaving the possibility that the behavior in some regimes is more complex. Using deterministic trajectories calculated in the slender body approximation, we are able to show that there is an appreciable dependence of $r_e$ on the angle $\psi$ that is not accounted for in Eq. \ref{rG}.

While the extent of the flow alignment of the helices appears to be relatively insensitive to the pitch, it would likely impact other important physical quantities. As the pitch gets very small, the tightly wound helix will approach a rigid cylindrical shell, which we expect would rotate with nearly complete fluid entrainment except at the ends. By contrast, if $p$ is large (compared to $r$), the amount of fluid displaced by the helix will be determined by filament length, with a logarithmic dependence on the filament thickness, and will be relatively insensitive to the helix radius. Finally, in this study we have only considered isolated helices, but interactions between helices, such as the nature of entanglements, will likely depend on $p/r$.

\begin{acknowledgments}
This work was supported by the AFOSR (FA9550-10-1-0473 and FA9550-14-1-0171). JSU is supported in 
part by the Georgetown Interdisciplinary Chair in Science Fund. We thank Peter Olmsted for helpful discussions.
\end{acknowledgments}

\newpage 
\bibliography{rost_helix_bib}

\end{document}